\documentclass[10pt, twocolumn]{IEEEtran}
\usepackage{setspace}
\usepackage{clrscode}
\usepackage{amsmath}
\usepackage{amssymb}
\usepackage[dvips]{graphicx}
\usepackage{epsfig}
\usepackage{hyperref}
\usepackage{bm}
\usepackage{subfigure}
\usepackage{balance}

\newcommand{\figsize}{0.22}

\newcommand{\E}{\mathbb{E}}

\newtheorem{theorem}{Theorem}

\newtheorem{remark}{Remark}

\begin{document}

\title{Age-Optimal Power Control for Status Update Systems with Packet Based Transmissions}
\author{
{Deli Qiao  and M. Cenk Gursoy}
\thanks{D. Qiao is with the School of Communication and Electronic Engineering, East China Normal University, Shanghai, China 200241 (email: dlqiao@ce.ecnu.edu.cn). M.C. Gursoy is with the Department of Electrical Engineering and Computer Science, Syracuse University, NY 13210 (email: mcgursoy@syr.edu)}
\thanks{This work has been supported in part by the National Natural Science Foundation of China (61671205).}}
\maketitle

\begin{abstract}
This paper investigates the average age of information (AoI) minimization in status update systems in which the update packets are transmitted with fixed rate. It is assumed that the source avoids queue-induced delay by generating and sending a new status update packet in each time slot. Assuming that the transmit power can be adapted based on the number of successive NACKs (negative acknowledgements) received, a closed-form expression for the average AoI is derived considering transmissions over block fading channel models. An optimization problem for minimizing the average AoI under average power constraints is formulated and a stochastic optimization algorithm for obtaining a feasible solution is proposed.  Numerical results on the proposed power control policy show that more power is preferred to be allocated to packets after a  certain number of failure packets in the low-power regime and the proposed power control reduces the average AoI compared to the constant power policy. An alternative on-off power control policy is also proposed and shown to achieve satisfactory performance.
\end{abstract}


\section{Introduction}

Timely update of information is critical in monitoring systems involving e.g., IoT/sensor networks with applications in forest fire alert systems, temperature and air/water pollution monitoring. With this motivation, age of information (AoI) has been introduced to measure the freshness of the status information in a remote system \cite{aoi}. Specifically, AoI is defined as the time elapsed since the generation of the last successfully received status information. As has been demonstrated, AoI is a  metric different from the existing ones measuring delay and latency and has attracted much interest recently (see, e.g., \cite{aoi-packetmanage}-\cite{aoi-harqmg1} and references therein). For instance, packet management policies, which preempt the old packets when a new packet arrives and  improve the system performance with respect to AoI, have been proposed and analyzed in \cite{aoi-packetmanage}. In \cite{aoi-update}, the optimal generation of the update packets have been investigated and efficient algorithms for determining the optimal update policy minimizing AoI have been designed. The optimal update policies for the energy harvesting nodes have been studied in \cite{aoi-ehmin}. Moreover, imposing a packet deadline has been shown to further reduce the average AoI \cite{aoi-deadline}.

\begin{figure}
    \centering
    \includegraphics[width=\figsize\textwidth]{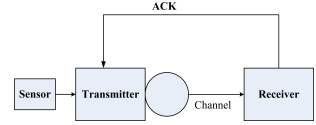}
    \caption{System model.}
    \label{fig:model}
\end{figure}

As exemplified above, most prior work is based on statistical modeling of the packet arrival and service time, where packets are generated and served randomly following certain distributions. Only few studies take into account the impact of wireless fading channels when such channels are used for the transmission of the update packets. Recently, in \cite{aoi-markov}, the authors have considered two-state Markov fading channels and derived closed-form expressions for the average AoI. In \cite{aoi-arq}, the authors have modeled the update packet delivery over block-fading channels at fixed transmission rates subject to constraints on the maximum number of transmission rounds,  and have quantified the average AoI and the tradeoff between the consumed energy and average AoI.
%
%

In this paper, we also consider the transmission of status packets  over fading channels at fixed rates. Instead of using fixed power, we assume that depending on the number of packets unsuccessfully received, the source allocates different amounts of power. For this general case, we derive the average AoI when transmission takes place over block fading channels. Based on the obtained average AoI, we formulate an optimization problem that minimizes the average AoI under average power constraints at the source. We design a stochastic optimization algorithm based on simulated annealing and evolutionary programming to obtain approximately optimal solutions.  We provide numerical results, identifying the performance improvements with power allocation.

The rest of this paper is organized as follows. Section II discusses the relevant preliminaries on system model, Markov chain modeling, and AoI. In Section III, the main results on the average AoI under the considered settings are presented. Numerical results are provided in Section IV, and Section V concludes this paper.

\section{Preliminaries}

\subsection{System Model}

Consider a remote system that provides status updates as shown in Fig. \ref{fig:model}. It is assumed that the status of the system is sensed and sent using status update packets (generated at the source)  to the destination over the wireless communication link between the source and the destination. Time is slotted into slots of duration $T$. We assume that an update packet of $R$ bits is generated every $T$ seconds, and transmitted over the wireless channel. We assume that no retransmission is needed. If a decoding failure occurs, the destination sends a NACK (negative acknowledgement) feedback to the source, which is a single-bit feedback. We assume that the NACK feedback can be received without error. On the other hand, for successful transmissions, there is no need for the destination to send ACK signal.  We consider a block fading channel in which the channel gain is assumed to be constant in each time slot and varies independently over different time slots.

\subsection{Markov Chain Model}

We assume no channel state information at the source, but the source knows the channel distribution information. Depending on number of consecutively received NACK signals, different transmission power levels are used. For instance, after receiving $m$ consecutive NACKs, the source sends the update packet with power $P_m$. The average power at the source is limited by $\bar{P}$.

Let us denote the number of consecutive NACKs as the state. State 0 denotes that the previous update packet is successfully transmitted, and state $m$ represents that $m$ consecutively sent packets have not been successfully received by the destination. With these state descriptions, we can model the transmissions of the status update packets as a discrete-time countable-state Markov process. 
Define $\epsilon_m$ as the decoding failure probability for the time slot after $m$ successive NACKs. Hence, the system enters state $m+1$ with probability $\epsilon_m, m=1,\ldots,M-1,\ldots$, while the system moves to state 0 with probability $1-\epsilon_m$.  Then, the state transition matrix is given by 
\begin{small}
\begin{align}\label{eq:stateprob}
\mathbf{A} = \left[
\begin{array}{llllll}
1-\epsilon_0 & \epsilon_0  & 0 & \cdots  & 0&\cdots\\
1-\epsilon_1 & 0  & \epsilon_1 &\cdots & 0 &\cdots\\
\vdots & \vdots  & \vdots & \ddots & \vdots & \vdots\\
1-\epsilon_{M-1} & 0  & 0 & \cdots  & \epsilon_{M-1}&\cdots\\
\vdots & \vdots  & \vdots & \ddots & \vdots & \vdots
\end{array}
\right]
\end{align}
\end{small}
whose $(i,j)^{th}$ component denotes the probability of state transition from state $i$ to state $j$.
The decoding error probability $\epsilon_m$ is defined as
\begin{small}
\begin{align}
\epsilon_m &= \Pr\{z[m]<z_m|z[m-1]<z_{m-1}\},\\
&= \Pr\left\{z[m]<\frac{2^{R}-1}{P_{m}}\right\}.\label{eq:epsm}
\end{align}
\end{small}
where $z[m]$ refers to the channel gain experienced by the update packet after $m$ successive failed packets, and (\ref{eq:epsm}) follows from the independent block fading assumption and the fact that decoding error occurs if $R > \log_2(1 + P_m z[m])$.

The steady state probability $\bm{\pi}=(\pi_0,\ldots,\pi_{M-1},\ldots)$ for the states can be derived from
\begin{align}
\bm{\pi} = \bm{\pi}\mathbf{A},\,\,\text{and}\,\,\sum_{j=1}^\infty \pi_j=1.
\end{align}
Now, the average power constraint can be expressed as follows:
\begin{small}
\begin{align}\label{eq:avgp}
\sum_{m=0}^{\infty}P_m\pi_m \le \bar{P}.
\end{align}
\end{small}

\subsection{Age of Information (AoI)}

\begin{figure}
\begin{center}
\subfigure[Example of age evolution.]{
    \label{fig:aoimodel} 
    \includegraphics[width=\figsize\textwidth]{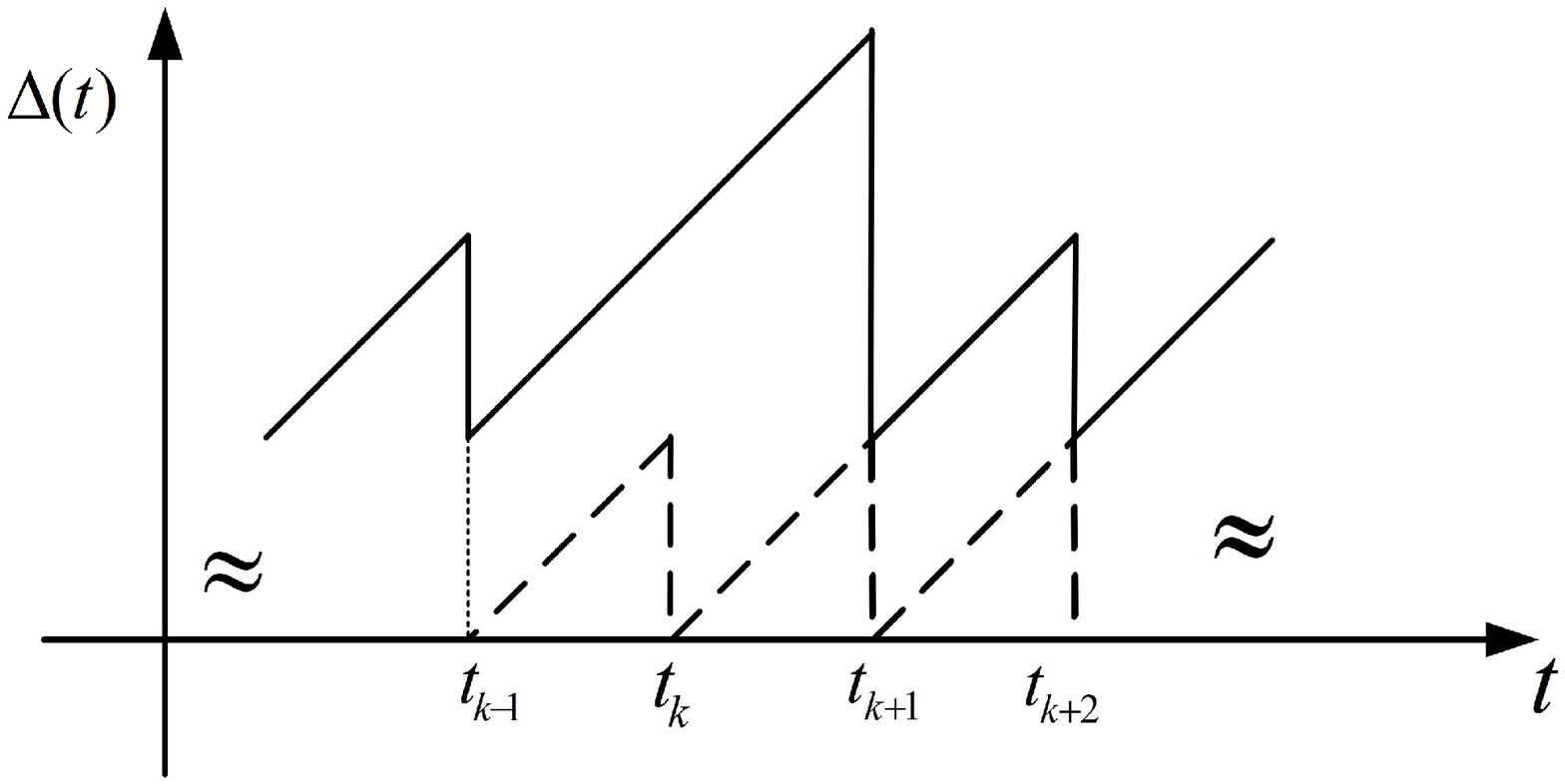}}
\subfigure[Reordered age evolution.]{
    \label{fig:aoimodel-rev} 
    \includegraphics[width=\figsize\textwidth]{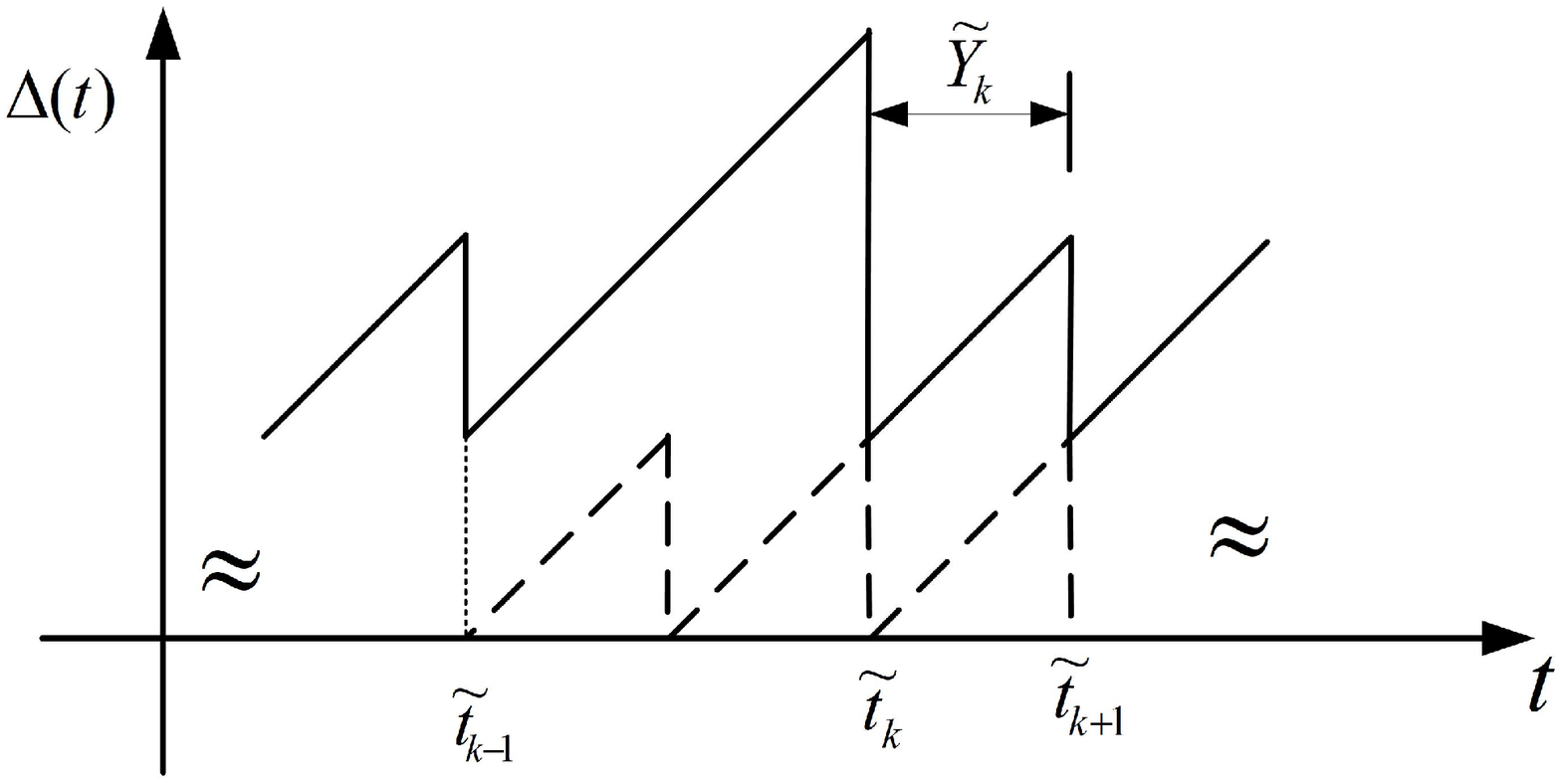}}
\end{center}\caption{Age of information (AoI).}
\end{figure}


AoI quantifies the freshness of the information about the status of a remote system. It is defined as the time elapsed since the generation of the last successfully received message containing system information. Assume that at any time $t$, the last successfully received packet was generated at time $U(t)$. Then, the age is defined as
\begin{align}
\Delta(t) = t- U(t).
\end{align}
Fig. \ref{fig:aoimodel} illustrates an example of the evolution of AoI in time, where $t_k$ represents the time when the $k$-th packet is generated. Note that the time slot is normalized as $T = 1$.  So $m$ is equivalent to the number of time slots for unsuccessful updates. After successfully receiving one packet, the age is reduced to one. Otherwise, the age continues to increase until another packet is successfully received. The average
AoI is given by
\begin{align}\label{eq:avgaoi}
\overline{\Delta} = \lim_{\tau\to\infty}\frac{1}{\tau}\int_0^\tau \Delta(t)dt.
\end{align}
In this paper, we assume that AoI can be measured in terms of time slots $T$. Please note that $T$ can take any suitable value. For instance, it can be determined by how frequently the status of the system is sensed. Without loss of generality, we assume $T=1$ in the following.

\section{Average AoI and Optimal Power Control}

Similar to the discussions in \cite{aoi-arq}, we can reindex the successfully received packets as shown in Fig. \ref{fig:aoimodel-rev}. In the following, the index $k$ stands for the $k$-th successfully received packet which may span over multiple unsuccessfully received packets. Define $\tilde{t}_k$ as the block index for the packet after the $k$-th successful update, $\tilde{Y}_k = \tilde{t}_{k+1}-\tilde{t}_k$ as the number of generated packets (time slots) between two consecutive successful updates. Note that $m=\tilde{Y_k}-1$ is the state for the $k$-th successful update that transits to state 0.

Similar to the conventional approaches for computing the average AoI, we divide the integral defined in (\ref{eq:avgaoi}) into trapezoids with area given by $\frac{\tilde{Y}_k(2+\tilde{Y}_k)}{2}$ such that
\begin{align}\label{eq:deltadef}
\overline{\Delta} = \lim_{k\to\infty}\frac{\sum_{i=1}^k\frac{\tilde{Y}_i(2+\tilde{Y}_i)}{2}}{\sum_{i=1}^k\tilde{Y}_i}=1+\frac{\E\{\tilde{Y}_k^2\}}{2\E\{\tilde{Y}_k\}}.
\end{align}
%
%

Then, we have the following characterization.
\begin{theorem}\label{theo:inddelta}
The average AoI for status update systems described above is given by
\begin{small}
\begin{align}\label{eq:inddeltafin}
\overline{\Delta} =\frac{3}{2}+\frac{\sum_{j=0}^\infty j\xi_j}{\sum_{j=0}^\infty \xi_j},
\end{align}
\end{small}
where
\begin{small}
\begin{align}\label{eq:xidef}
\xi_0=1,\,\,\,\xi_j&=\prod_{m=0}^{j-1}\epsilon_m,\,\,j=1,2,\ldots.
\end{align}
\end{small}
\end{theorem}
\emph{Proof:} Please see Appendix \ref{app:inddelta} for details.\hfill$\square$
\begin{remark}
Using the above result, we can easily show that $\overline{\Delta}$ is lowerbounded by 1.5 since $\sum_{j=0}^\infty \xi_j\ge1$ and $\sum_{j=0}^\infty j\xi_j\ge0$.
\end{remark}

\begin{remark}
In case of no power control, we can see that $\epsilon_0=\epsilon_1=\ldots=p=\Pr\left\{z[m]<\frac{2^{R}-1}{\bar{P}}\right\}$. Then, we have $\xi_j=p^{j}$, $\sum_{j=0}^{\infty}\xi_j=\frac{1}{1-p}$, and $\sum_{j=0}^{\infty}j\xi_j=\frac{p}{(1-p)^2}$. Therefore, in this case, $\overline{\Delta}=\frac{3-2p}{2(1-p)}$.
\end{remark}

In the following, we analyze the optimal power control scheme that minimizes the average AoI. We can formulate the optimization problem under the average power constraints as follows.
\begin{align}
\min_{\mathbf{P}}\,& \,\overline{\Delta},\hspace{.5cm}\text{s.t.}\,\, \sum_{m=0}^\infty P_m\pi_m \le \bar{P}.
\end{align}
Note that the objective function $\overline{\Delta}$ depends on the power allocation vector $\mathbf{P} = (P_0, P_1, \ldots, P_{M-1},\ldots)$ through $\bm{\epsilon}=(\epsilon_0,\ldots,\epsilon_{M-1},\ldots)$, and hence is non-convex in $\mathbf{P}$. Therefore, the optimization problem is also non-convex. We first consider an on-off type control policy, in which the transmitter keeps silent for the first $\tau$ states, and then sends packets with constant power level $P_\tau$, i.e., $\mathbf{P}_\tau=(\underbrace{0,\ldots,0}_{\tau},P_\tau,\ldots,P_\tau)$. Note that given $\tau$, we can determine the value of $P_\tau$ that satisfies the average power constraints by bi-section method, and the optimal $\tau$ can be selected from a limited range to minimize the average AoI.

In the following, we resort to stochastic optimization techniques to obtain approximately optimal solutions. Specifically, we modify the simulated annealing algorithm as used in \cite{powerarq} by generating a new point following Cauchy distribution. As shown in \cite{fep}, this kind of evolutionary programming algorithm can speed up the convergence. We start from the point of $\bm{\epsilon}$ associated with the on-off power control policy $\mathbf{P}_\tau$, and then generate a random set of $\bm{\epsilon}'$ with deviations from $\bm{\epsilon}$ given by a standard Cauchy random variable multiplied by the current temperature $T_n$ so that the convergence to the optimal solution can be faster, i.e.,
\begin{align}
\bm{\epsilon}' = \bm{\epsilon} + T_n x,\label{eq:neweps}
\end{align}
where the probability density function of $x$ is given by $f(x) = \frac{1}{\pi}\frac{1}{1+x^2}$. Note that the temperature denotes a numerical value that controls the muting process. In numerical evaluations, we adopt a typical temperature change given by \cite{sa}
\begin{align}
T_n= \frac{T_0}{n},
\end{align}
where $T_0$ is the initial temperature. Subsequently, we can derive the associated transmit power $\hat{\mathbf{P}}$ from (\ref{eq:epsm}) and obtain the average power according to the left-hand-side of (\ref{eq:avgp}). If the average power constraint (\ref{eq:avgp}) is satisfied, we evaluate the average AoI from (\ref{eq:inddeltafin}).  Note that the set of $\bm{\epsilon}'$ may be viewed as a better one with a certain temperature-dependent probability even if the resulting average AoI is higher. This process is called muting and enables the system to avoid local minima for non-convex optimization problems. When the temperature is very high, the muting occurs very frequently and becomes less often as the temperature decreases gradually. To summarize, we propose Algorithm 1 below to obtain an approximately optimal solution. Note that once given the channel statistics, the algorithm can be performed offline. Additionally, computational complexity in each iteration is low due to the fact that in each iteration a random set of transition probabilities is generated, and relatively simple computations (e.g., the computation of the average age using (\ref{eq:inddeltafin}) and checks are performed. We would like to note that $M$ in the algorithm is some large value representing the largest state value for the Markov chain. In this way, we have a finite state Markov chain approximation to the problem. Obviously, when $M$ goes to infinity, the resultant policy will converge to that of the original problem.


\begin{codebox}
\Procname{$\proc{Algorithm 1: Proposed Power Control}$}
 \li      \textbf{Input:} $(M,\bar{P},R)$;
 \li      Initialize $N$, $T_n = T_0$, $T_{\min}$, $\mathbf{P}^*=\mathbf{P}_\tau$, \\
 obtain $\bm{\epsilon}=(\epsilon_1,\ldots,\epsilon_M)$ with $\mathbf{P}^*$, \\and compute $\overline{\Delta}_{0}$ from (\ref{eq:inddeltafin});
 \li  Let $\overline{\Delta}_{\min}=\overline{\Delta}_{0}$;
 \li      \While $T_n\ge T_{\min}$
 \li \Do    $T_n = \frac{T_0}{n}$;
 \li  \For $i=1$ \To $N$
 \li  \Do      Generate a random $\mathbf{A}$ with $\bm{\epsilon}'$ in (\ref{eq:neweps});
 \li         Compute the average power $\hat{\mathbf{P}}$ and $\hat{\bar{P}}$ from $\mathbf{A}$;
 \li        \If  $\hat{\bar{P}}<\bar{P}$
 \li \Do Compute $\hat{\overline{\Delta}}$ from $\mathbf{A}$;
 \li    Generate a random number $s\in(0,1)$;
 \li  \If $s<e^{-\frac{\hat{\overline{\Delta}}-\overline{\Delta}_0}{T_n}}$
 \li  \Do $\overline{\Delta}_0=\hat{\overline{\Delta}}$, $\bm{\epsilon}=\bm{\epsilon}'$;
 \li \If $\hat{\overline{\Delta}}\le \overline{\Delta}_{\min}$
 \li   \Do $\overline{\Delta}_{\min} = \hat{\overline{\Delta}}$;
 \li  $\mathbf{P}^* = \hat{\mathbf{P}}$;\End\End
 \li \Else
 \li Solution is not feasible;\End\End\End
 \li \textbf{Output:} $\mathbf{P}^*, \overline{\Delta}_{\min}$;
\end{codebox}


\section{Numerical Results}

\begin{figure}
\begin{center}
\subfigure[The average AoI as a function of $\bar{P}$.]{
    \label{fig:aoiinsnr} 
    \includegraphics[width=\figsize\textwidth]{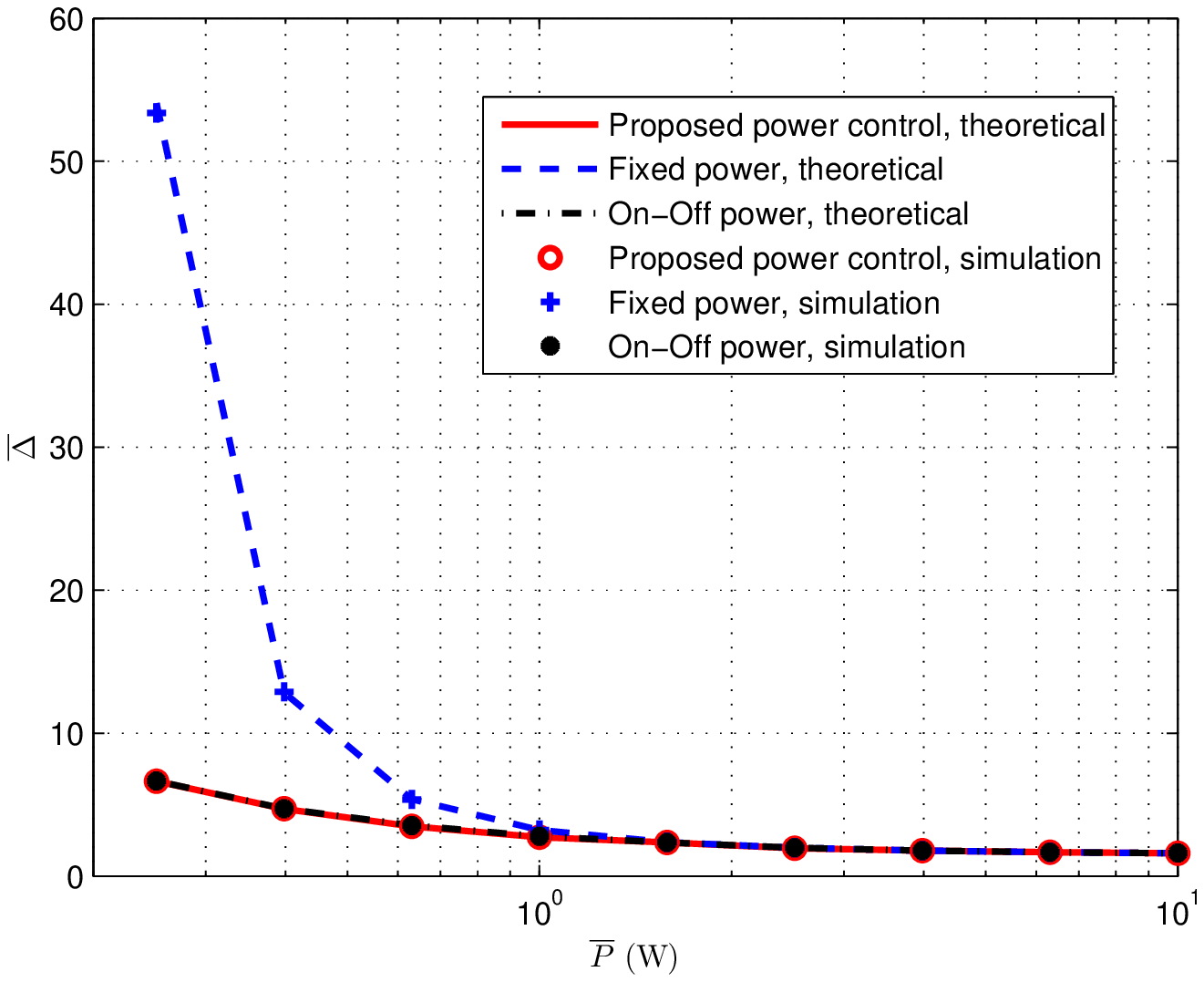}}
\subfigure[Power allocation as state $m$ varies when $\bar{P}=10^{-0.6}$ W.]{
    \label{fig:allopowerinsnr} 
    \includegraphics[width=\figsize\textwidth]{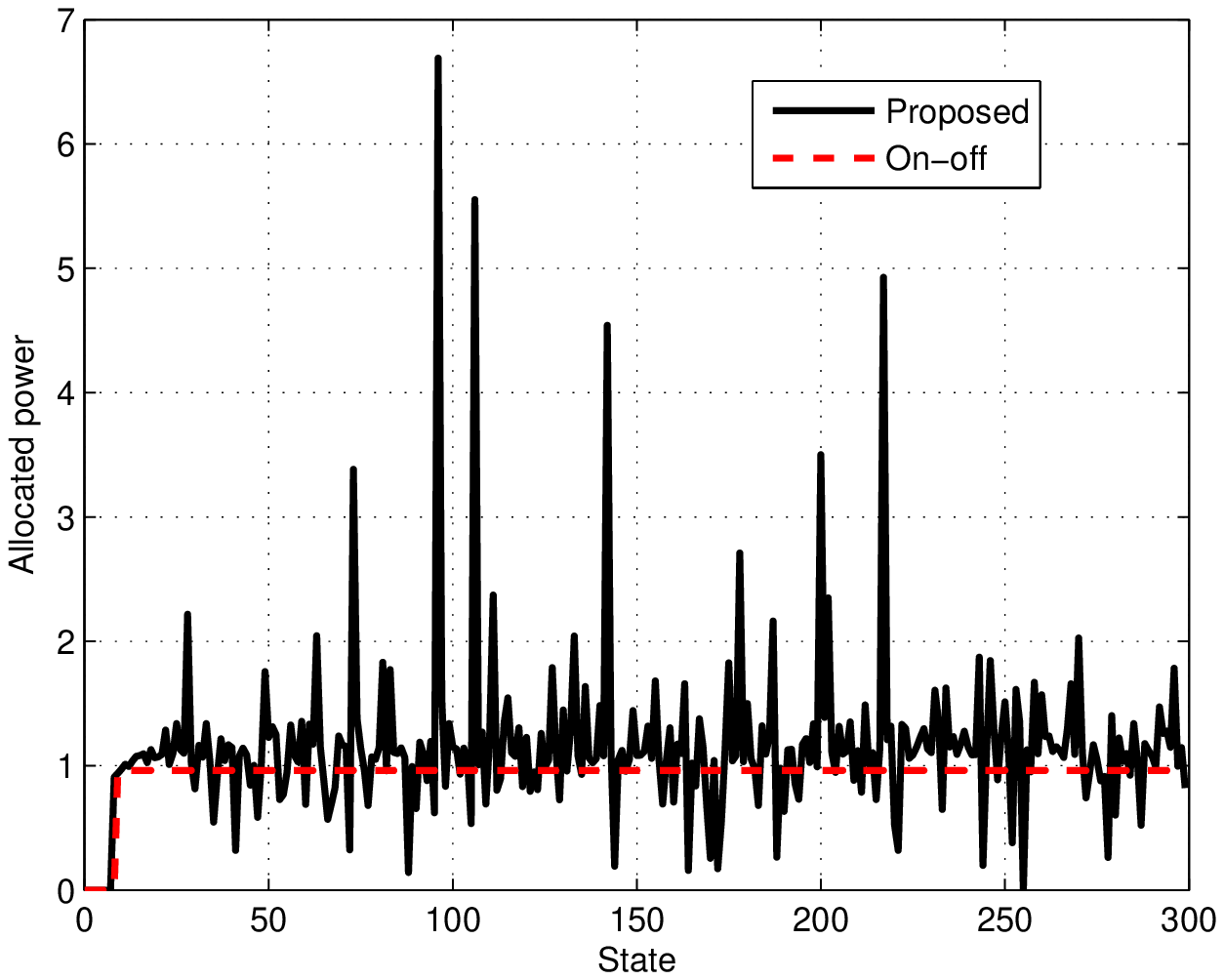}}
\end{center}\caption{Average AoI and allocated power in state.}
\end{figure}
%
%

In the following, we consider Rayleigh fading channels with unit mean. We generate $10^6$ independent channels and compute the average of the AoI for simulation. We set $T_0=1$.

In Fig. \ref{fig:aoiinsnr}, we plot the average AoI as a function of $\bar{P}$ for the block fading channels. We assume $R=1$. In the simulation, we assume that the largest state value is $M=300$ to approximate the countable-state Markov chain. 
We can first see that the theoretical results match the simulation results, verifying the theoretical analysis. Also, we can see that with the proposed power control algorithm, the average AoI of the status update system can be reduced significantly, especially when $\bar{P}$ is low. For instance, when $\bar{P}=10^{-0.6}$ W, the average AoI can be reduced by more than $80\%$ with the proposed power control policy. On the other hand, when the average power $\bar{P}$ is large, it can be expected that constant power achieves performance close to the proposed power control policy. Moreover, both policies can achieve average AoI close to the lowerbound of 1.5 when $\bar{P}$ is high. Also, the on-off power control policy achieves performance close to the proposed one justifying the efficiency of the on-off power control policy. In Fig. \ref{fig:allopowerinsnr}, we plot the associated power control policies as state $m$ varies when $\bar{P}=10^{-0.6}$ W. From the figure, we can find that there is a sudden jump of the allocated power starting from state 7. In other words, if the transmission fails after the seventh consecutively failed transmissions, there will be an increase in the transmit power. This tells us that in the low power regime, more power is allocated for the larger state to reduce the decoding failure probabilities so that the packet can be more likely to be successfully received. In this way, the average AoI can be dramatically reduced.
%
%
%
%
%
%
%

\section{Conclusion}

In this paper, we have considered the power allocation problem for minimizing the AoI in status update systems. We have assumed that the fixed size update packets are generated and transmitted over wireless block-fading channels.  We have proposed a power allocation policy based on the number of successive NACKs received. By deriving a closed-form expression for the average AoI in block fading channels, we have formulated the AoI minimization problem subject to average power constraints. We have resorted to stochastic optimization techniques and proposed a simulated annealing and evolutionary programming based algorithm for power control. Through numerical results, we have found that the proposed power control policy can reduce the average AoI significantly in the low power regime. We have also proposed an on-off power control policy, where the transmit power level is zero for small ages and constant for larger ages, and shown its efficiency in reducing the average AoI in the low power regime. 

\appendix
\subsection{Proof of Theorem \ref{theo:inddelta}}\label{app:inddelta}

Suppose that the $k$-th successful status update spans over $m-1$ unsuccessful update packets. Then, we know that the Markov chain transits from state $m-1$ to state $0$ when transmitting the $m$-th update packet. Note that $\tilde{Y}_k=m$. Therefore, we can write
\begin{small}
\begin{align}
\Pr\{\tilde{Y}_k=1\} &=1-\epsilon_0,\,
\Pr\{\tilde{Y}_k=m\} = \prod_{i=0}^{m-2}\epsilon_i(1-\epsilon_{m-1}),\,\forall m\ge 2.
\end{align}
\end{small}
Next, we can show that
\begin{small}
\begin{align}
\E\{\tilde{Y}_k\} &= \sum_{m=1}^\infty m\Pr\{\tilde{Y}_k=m\}\\
&= 1-\epsilon_0+\sum_{m=2}^\infty m \prod_{i=0}^{m-1}\epsilon_i(1-\epsilon_{m-1})\label{eq:indproof-0}\\
& = \sum_{j=0}^\infty\xi_j,\label{eq:indX1}
\end{align}
\end{small}
where $\xi_j$ defined in (\ref{eq:xidef}) is incorporated. Also,
\begin{small}
\begin{align}
\E\{\tilde{Y}^2_k\}& = \sum_{m=1}^\infty m^2\Pr\{\tilde{Y}_k=m\}\\
&=1-\epsilon_0+\sum_{m=2}^\infty m^2 \prod_{i=0}^{m-1}\epsilon_i(1-\epsilon_{m-1})\label{eq:indproof-1}\\
& = 2\sum_{j=0}^{\infty}j\xi_j+ \sum_{j=0}^{\infty}\xi_j\label{eq:indX2}
\end{align}
\end{small}
Substituting (\ref{eq:indX1}) and (\ref{eq:indX2}) into (\ref{eq:deltadef}) gives us (\ref{eq:inddeltafin}).\hfill$\square$

\balance


\begin{thebibliography}{99}
\bibitem{aoi} S. Kaul, R. Yates, M. Gruteser, Real-time status: How often should one update?, in: Proc. IEEE International Conference on Computer Communications (INFOCOM), 2012. doi:10.1109/INFCOM.2012.6195689.

\bibitem{aoi-packetmanage} M. Costa, M. Codreanu, and A. Ephremides, ``On the age of information in status update systems with packet management,'' \emph{IEEE Trans. Inf. Theory}, vol. 62, no. 4, pp. 1897 - 1910, Apr. 2016.

\bibitem{aoi-update} Y. Sun \emph{et al.}, ``Update or wait: how to keep your data fresh,'' \emph{IEEE Trans. Inf. Theory}, vol. 63, no. 11, pp. 7492-7508, Nov. 2017.



\bibitem{aoi-ehmin} X. Wu, J. Yang, and J. Wu, ``Optimal status update for age of information minimization with an energy harvesting source,'' \emph{IEEE Trans. Green Commun. Netw.}, vol. 2, no. 1, pp. 193-204, Mar. 2018.

\bibitem{aoi-preempt} S. Farazi, A. G. Klein, and D. R. Brown, ``Age of information in energy harvesting status update systems: when to preempt in service?'' in 2018 IEEE Internation Symposium on Information Theory (ISIT), pp. 2436-2440, Vail, Colorado, June 2018.

\bibitem{aoi-deadline} C. Kam \emph{et al.}, ``On the age of information with packet deadlines,'' \emph{IEEE Trans. Inf. Theory}, vol. 64, no. 9, pp. 6419-6428, Sep. 2018.

\bibitem{aoi-harq} E. Ceran, D. Gunduz, and A. Gyorgy, ``Average age of information with hybrid ARQ under a resource constraint,'' IEEE WCNC 2018.

\bibitem{aoi-mac} H. Sac, T. Bacinoglu, E. Uysal-Biyikoglu, and G. Durisi, ``Age-optimal channel coding blocklength for an M/G/1 queue with HARQ,'' IEEE SPAWC 2018.

\bibitem{aoi-harqmg1} H. Sac, T. Bacinoglu, E. Uysal-Biyikoglu, and G. Durisi, ``Age-optimal channel coding blocklength for an M/G/1 queue with HARQ,'' IEEE SPAWC 2018.

\bibitem{aoi-markov} L. Huang and L. Qian, ``Age of information for transmissions over Markov channels,'' in 2017 IEEE Global Communications Conference (Globecom), Singapore, Dec. 2017.

\bibitem{aoi-arq} J. Gong, X. Chen, and X. Ma, ``Energy-age tradeoff in status update communication systems with retransmission,'' to appear at the IEEE Globecom 2018. Available: \url{https://arxiv.org/abs/1808.01720}.
		
\bibitem{powerarq} M. M. Butt, E. A. Jorswieck, and N. Marchetti, ``On optimizing power allocation for reliable communication over fading channels with uninformed transmitter,'' \emph{IEEE Trans. Wireless Commun.}, vol. pp, no. 99, pp. 1-1, 2018.
		
\bibitem{sa} H. Szu and R. Hartley, ``Fast simulated annealing,'' \emph{Physics Letters A}, vol. 122, no. 3-4, pp. 157¨C162, June 1987.

\bibitem{fep} X. Yao, Y. Liu, and G. Lin, ``Evolutionary programming made faster,'' \emph{IEEE Trans. Evolution. Comp.}, vol. 3, no. 2, pp. 82-102, July 1999.



	\end{thebibliography}
\end{document}